\documentclass{svjour3}                     
\smartqed  
\usepackage{graphicx}
\usepackage{mathptmx}      
\usepackage{natbib}
\newcommand {\gtsim} {\ {\raise-.5ex\hbox{$\buildrel>\over\sim$}}\ }
\newcommand {\ltsim} {\ {\raise-.5ex\hbox{$\buildrel<\over\sim$}}\ }
\newcommand{\commentpcf}[1]{}

\newcommand{\muG}{$\mu $G}
\newcommand{\microns}{$\mu$m}
\newcommand{\microG}{$\mu$G}

\newcommand{\glong}{$\ell$}
\newcommand{\glat}{$b$}

\newcommand{\OI}{O$^{\rm o}$}
\newcommand{\OII}{O$^{\rm +}$}

\newcommand{\NHI}{$N$(H$^{\rm o}$)}

\newcommand{\HI}{H$^{\rm o}$}
\newcommand{\TiII}{Ti$^{\rm +}$}
\newcommand{\DI}{D$^{\rm o}$}

\newcommand{\CaII}{Ca$^{\rm +}$}
\newcommand{\CaIII}{Ca$^{\rm ++}$}
\newcommand{\CII}{C$^{\rm ++}$}
\newcommand{\MgII}{Mg$^{\rm +}$}
\newcommand{\MgI}{Mg$^{\rm o}$}
\newcommand{\HII}{H$^{\rm +}$}
\newcommand{\CIIstar}{C$^{ + *}$}

\newcommand{\NeIII}{Ne$^{\rm ++}$}
\newcommand{\NeII}{Ne$^{\rm +}$}
\newcommand{\NeI}{Ne$^\circ$}
\newcommand{\ArI}{Ar$^\circ$}

\newcommand{\NII}{N$^+$}
\newcommand{\SII}{S$^+$}

\newcommand{\HeI}{He$^{\rm o}$}

\newcommand{\nHI}{$n \mathrm{(H^\circ)}$}
\newcommand{\nHII}{$n \mathrm{(H^+)}$}
\newcommand{\nHeI}{$n \mathrm{(He^\circ)}$}

\newcommand{\kms}{\hbox{km s$^{-1}$}}
\newcommand{\deeg}{\hbox{$^{\rm o}$}}

\newcommand{\lya}{\hbox{Ly$\alpha$}}

\newcommand{\Halpha}{H$\alpha$}
\newcommand{\cmtwo}{cm$^{-2}$}

\newcommand{\cc}{cm$^{-3}$}

\newcommand\araa{{ARA\&A}}%
\newcommand\apj{{ApJ}}%
\newcommand\apjl{{ApJ}}%
\newcommand\apjs{{ApJS}}%
\newcommand\aap{{A\&A}}%
\newcommand\aaps{{A\&AS}}%
\newcommand\ssr{{Space~Sci.~Rev.}}%
\newcommand\nat{{Nature}}%
\newcommand\jgr{{J.~Geophys.~Res.}}%
\def\eps@scaling{1.0}%
\newcommand\epsscale[1]{\gdef\eps@scaling{#1}}%
\newcommand\plotone[1]{%
 \typeout{Plotone included the file #1}
 \centering
 \leavevmode
 \includegraphics[width={\eps@scaling\columnwidth}]{#1}%
}%
\newcommand\plottwo[2]{{%
 \typeout{Plottwo included the files #1 #2}
 \centering
 \leavevmode
 \columnwidth=.50\columnwidth
 \includegraphics[width={\eps@scaling\columnwidth}]{#1}%
 \hfil
 \includegraphics[width={\eps@scaling\columnwidth}]{#2}%
}}%
\journalname{SSRv}
\begin{document}

\title{Is the Sun Embedded in a Typical Interstellar Cloud? }
\subtitle{Connecting Interstellar Gas in and out of the Heliosphere}

\titlerunning{Is the Sun Embedded in a Typical Interstellar Cloud?}        

\author{Priscilla C. Frisch }

\institute{P. Frisch \at  University of Chicago, \\
	      Dept. of Astronomy and Astrophysics, \\
              \email{frisch@oddjob.uchicago.edu}  }         

\date{Received: date / Accepted: date}

\maketitle

\begin{abstract}
The physical properties and kinematics of the
partially ionized interstellar material (ISM) near the Sun are typical
of warm diffuse clouds in the solar vicinity.  The direction of the
interstellar magnetic field at the heliosphere, the polarization of
light from nearby stars, and the kinematics of nearby clouds are
naturally explained in terms of the S1 superbubble shell.  The
interstellar radiation field at the Sun appears to be harder than the
field ionizing ambient diffuse gas, which may be a consequence of the
low opacity of the tiny cloud surrounding the heliosphere.
The spatial context of the Local Bubble is consistent with our location
in the Orion spur.

\end{abstract}

\section{Introduction} \label{intro}

Observations of interstellar gas in the Milky Way Galaxy span over ten
orders-of-magnitude in spatial scales, and over six
orders-of-magnitude in temperature.  Interstellar material (ISM) is
observed at the Earth's orbit, where interstellar \HeI\ has been
counted by the Ulysses GAS detector and is detected through
fluorescence of solar 584 A
\citep{WellerMeier:1981,FlynnVallerga:1998,Salernoetal:2003He,Witte:2004,Moebiusetal:2004}.
Interstellar \HI\ and other neutral interstellar atoms are driven into
the heliosphere by the relative Sun-cloud motion of 26.2 \kms, where
they become ionized and processed into pickup ions and anomalous
cosmic rays.  The question arises: 'Is the interstellar cloud feeding
gas and dust into the heliosphere a typical interstellar cloud?'  We
know more about the circumheliospheric interstellar material (CHISM)
than other clouds, because theoretical models combine $in~situ$ data
with observations of nearby stars to model the cloud opacity
profile. In this paper, the kinematics, temperature, ionization,
composition and density of the CHISM are compared to low density ISM
seen in the solar neighborhood.  The answer to the title question is
'yes'.

The properties of the heliosphere are governed by the Local Bubble
 (LB) void and the Local Interstellar Cloud (LIC).  The LB is
 transparent to radiation, exposing the LIC to a diffuse interstellar
 radiation field that includes both a soft X-ray background that
 anticorrelates with the \HI\ column density \citep{Bowyeretal:1968},
 and radiation from distant hot and white dwarf stars
 \citep{Gondhalekaretal:1980,Vallerga:1998}.  The LIC and ISM
 surrounding the Sun are part of the shell of a superbubble expanding
 into the low density interior of the LB.  The LIC ionization caused
 by the radiation field of the hot stars bordering the LB then
 dominates the heliosphere boundary conditions, while the ram pressure
 of gas in the expanding superbubble configures the heliosphere into
 the familiar raindrop shape.

\section{Galactic Environment of the Sun}

By now it is well known that the Sun resides in a region of space with
very low average interstellar densities, the Local Bubble, formed by
ISM associated with the ring of young stars bounding this void and
known as Gould's Belt \citep{Frisch:1995}.  The missing part of our
understanding has been the origin of the Gould's Belt stars, or the
Orion spur containing these stars.  The Orion spur, which is located
on the leading edge or convex side of the Sagittarius arm, is not
included in models of the Milky Way spiral arms.  However recent
advances in our understanding of the formation of spurs, or
'feathers', on spiral arms naturally explains the origin of the
Gould's Belt stars and the Orion spur.  The interaction between the
gaseous disk and the gravitational potential of spiral arms in the
presence of a magnetic field induces self-gravitating perturbations
that develop into two-dimensional flows that become unstable and
fragment, driving spurs, or 'feathers', of star-forming material into
low-density interarm regions (e.g. \citep{ShettyOstriker:2006}).  The
Orion spur can be seen extending between galactic longitude $\ell \sim
60^\circ$ distance $\sim 1.2$ kpc, and $\ell \sim 170^\circ$ distance
$\sim 0.7$ kpc, in Fig. 10 in \citep{Lucke:1978}, together with two
other spurs.  These three spurs have pitch angles with respect to a
circle around the galactic center of $\sim 40^\circ - 55^\circ $, and
are separated by 550--1500 pc, values that are normal for MHD models
of spur formation \citep{ShettyOstriker:2006}.  Star formation occurs
in spurs, and the Gould's Belt stars can thus naturally be explained
by the Orion spur of which they are part.  The Local Bubble is at the
inner edge of the Orion spur.  Based on MHD models of spur formation,
we must conclude that the Gould's Belt environment of the Sun is
normal, answering the question posed above with a 'yes'.

\section{Heliospheric ISM}\label{sec:inside}

The CHISM ionization balance is governed by photoionization and
recombination, so that neutral atoms and dust in the heliosphere trace
the cloud physics, as well as the composition, temperature, density,
and origin (Slavin and Frisch \citep{SlavinFrisch:2008}).  $In~situ$
gas and dust data, combined with radiative transfer models of CHISM
ionization, test the "missing-mass" premise that assumes the combined
interstellar atoms in gas and dust provide an invariant tracer of the
chemical composition of the ISM
\citep{SavageSembach:1996araa,Frischetal:1999,SlavinFrisch:2008}.
This test is potentially interesting because Gruen and Landgraf
\citep{GruenLandgraf:2000} have shown that large and small dust grains
couple to interstellar gas over different spatial scales, so that in
the presence of active or recent grain shattering by interstellar
shocks, local and global values for the gas-to-dust mass ratio may
differ.

Interstellar particles with gyroradii larger than the distance between
the particle and heliopause typically penetrate the heliosphere.  If
thermal and magnetic pressures are equal in the CHISM, then the
magnetic field strength is $\sim 2.7~\mu \mathrm{G}$ (Slavin and
Frisch \citep{SlavinFrisch:2008}).  Depending on the strength of the
radiation field responsible for grain charging through photoejection
of electrons, interstellar dust grains entering the heliosphere have
radii larger than $\sim 0.06 - 0.2$ \microns\
\citep{KruegerGruen:2008,Frischetal:1999,CzechowskiMann:2003b}.
Grains with radii $\sim 0.01 - 0.09$ \microns\ traverse the bow shock
region, but are deflected around the heliopause with other charged
populations.  When STARDUST observations of interstellar grains become
available, it will be possible to verify the missing-mass premise that
the composition of the CHISM consists of the sum of elements in the
gas and dust phases, and check whether solar abundances apply to the
CHISM.  If the missing-mass assumption is wrong it would explain the
$\sim 50$\% difference between the gas-to-dust mass ratios found from
$in~situ$ observations of interstellar grains, and missing mass
arguments utilizing radiative transfer models
\citep{SlavinFrisch:2008}.

Only the most abundant interstellar elements with first ionization
potentials $\gtsim 13.6$ eV are observed in the heliosphere in
detectable quantities, including H, He, N, O, Ne, and Ar.  Each of
these elements is observed in at least two forms, pickup ions (PUI)
and anomalous cosmic rays (ACR).  ACRs are accelerated PUIs.  Pickup
ions are formed when interstellar neutrals become ionized through
either charge-exchange with the solar wind (H, N, O, Ne, Ar),
photoionization (He, H), or electron impact ionization (He, N, Ar).
Helium data yield the best temperature, He density, and velocity data,
since the He charge-exchange cross-section with the solar wind is low
and He penetrates to within $\sim 0.5$ au of the Sun before ionization
by photons and electron impact become significant
\citep{Moebiusetal:2004}.  The He data indicate for the CHISM: $T =
6300 \pm 400$ K, \nHeI=$0.015 \pm 0.002$ \cc, and $V =-26.2 \pm 0.5$
\kms, and an upwind direction of $\lambda = 255.0^\circ \pm
0.6^\circ$, $\beta = 5.2^\circ \pm 0.3^\circ$ (corrected for J2000
coordinates, \citep{Moebiusetal:2004,Witte:2004}).  Early December
each year the Earth passes through a cone of gravitationally focused
He, extending over 5 au downwind of the Sun \cite{McComasetal:2004Hshadow}.

Hydrogen is the most abundant ISM observed in the heliosphere, however
the initial thermal interstellar velocity distribution of \HI\ is
modified and deformed as \HI\ enters and propagates through the
heliosphere.  Interpretation of the \lya\ fluorescence and PUI data
require corrections for the weak coupling between \HI\ and the
interstellar magnetic field outside of the heliosphere due to
\HI-\HII\ interactions, strong filtration through charge-exchange
between interstellar protons and \HI\ in the heliosheath regions,
deformation of the \HI\ velocity distribution as H atoms enter and
propagate through the heliosphere, and the solar-cycle dependent
variation in the ratio of radiation pressure and gravitational forces.
These effects are discussed elsewhere in this volume (e.g. Bzowski,
Quemerais, Wood, Opher, Pogorelov).  These observations and models of
PUI H inside of the heliosphere are consistent with an H density at
the termination shock of $\sim 0.11 $ \cc, and when combined with
filtration values yield an interstellar density of
$n(\mathrm{H^\circ}) \sim 0.195 \pm 0.02 $ \cc\ for the CHISM
\citep{Bzowskietal:2007}.  The H filtration factor is based on the
Moscow Monte Carlo model, which also yields a CHISM plasma density of
\nHII$=0.04 \pm 0.02$ \cc.  These results are in excellent agreement
with the completely independent radiative transfer results that
conclude \nHI$=0.19 - 0.20 $ \cc, and \nHII$=0.07 \pm 0.02$ \cc\ for
the CHISM \citep{SlavinFrisch:2008}.

Comparisons between abundances of neutrals in the CHISM as predicted
by radiative transfer studies, with interstellar neutral abundances
based on PUI and ACR densities corrected to values at the termination
shock, require that the filtration of neutrals crossing outer
heliosheath regions is understood
\citep{CummingsStone:2002,MuellerZankfilt:2004,Izmodenovetal:1999}.
In the heliosheath regions, charge-exchange with interstellar protons
increases filtration of O, and reverse charge-exchange potentially
allows interstellar \OII\ into the heliosphere.  Electron impact
ionization contributes to filtration of N and Ar.  Reverse
charge-exchange between interstellar ions and protons in the outer
heliosheath is insignificant for all elements except possibly O and H.
The range of filtration factors found for H, He, N, O, Ne, and Ar are
listed in \citep{SlavinFrisch:2008}.  Correcting PUI densities given
by \citep{GloecklerFisk:2007} for the termination shock with
calculated filtration factors yields interstellar densities for these
elements that are consistent, to within uncertainties, with the range
of neutral densities predicted by the CHISM radiative transfer models
for these elements (Table 4 in \citep{SlavinFrisch:2008}).  In the
radiative transfer models, abundances of H, He, N and O are variables
that are required to match the data, but Ne and Ar abundances are
assumed at 123 ppm and 2.82 ppm, respectively.

The only measurements of Ne in the local ISM are the $in~situ$ PUI and
ACR data; Ne is a sensitive tracer of the ionization conditions of the
CHISM because three ionization states are present in significant
quantities, \NeI:\NeII:\NeIII$\sim$1.0:3.3:0.8 (Slavin and Frisch
\citep{SlavinFrisch:2008}).  The radiative transfer model results are
also consistent with the Ne abundance of $\sim 100$ ppm found in the
Orion nebula \citep{Simpson:2004}, and within the range of
uncertainties for the solar Ne abundance.

Neutral Ar traces the equilibrium status of the CHISM because \ArI\
and \HI\ are the end products of processes with similar recombination
rates, but have different photoionization rates (Slavin, this volume).
The radiative transfer models \citep{SlavinFrisch:2008} together with
the PUI Ar data indicate the CHISM is in ionization equilibrium.  The
ratio \ArI/\HI$\sim 1.0 \times 10^{-6}$ in the CHISM found from PUI
data and radiative transfer models is comparable to interstellar
values towards nearby stars based on the FUV data, \ArI/\HI$\sim 1.2
\times 10^{-6}$ \citep{Jenkinsetal:2000}.  Agreement with the FUSE
data can be achieved by a small increase in the assumed Ar abundances
in the \citep{SlavinFrisch:2008}.

Isotopes in PUIs, ACRs, and He indicate that the CHISM is formed from
similar material as the Sun.  The ratios $^{22}$Ne/$^{20}$Ne $\sim
0.073$ and $^{18}$O/$^{16}$O $\sim 0.002$ are close to isotopic ratios
in the solar wind \citep{CummingsStone:2007,Leske:2000}.  He data
gives $^{3}$He/$^{4}$He $\sim 1.7 - 2.2  \times 10^{-4}$, which is similar to
meteoritic and HII region values
\citep{Salernoetal:2003He,GloecklerFisk:2007}.  Evidently the expected
$^{3}$He enrichment of the ISM by nucleosynthesis in low-mass stars
has not affected the CHISM.  The $^{22}$Ne isotope indicates that the
CHISM is not significantly mixed with ejecta from Wolf Rayet stars
common to OB associations, where $^{22}$Ne would be enriched by
He-burning.  The CHISM gas therefore appears isotopically similar to
solar system material, and $^{3}$He values are consistent with
isotopic ratios in HII regions.

Summarizing, observations of interstellar products inside of the
heliosphere yield densities and abundances for H, He, N, O, Ne, and Ar
that are in agreement with radiative transfer models of LIC absorption
components in the star $\epsilon$ CMa.  Argon has similar abundances,
\ArI/\HI, in the CHISM and towards near white dwarf stars.  Isotopic
ratios suggests that the CHISM has a solar composition.  $In~situ$
observations of interstellar dust grains yield a gas-to-dust mass
ratio that varies by 50\% or more from values predicted by radiative
transfer models, indicating that the either the abundances of elements
depleted onto dust grains or the true metallicity of the CHISM is not
understood.  The CHISM abundances determined from $in~situ$ data are
consistent with abundances typical of low density ISM, so that based
on $in~situ$ observations of ISM we conclude that the answer posed
above is 'yes'.

\section{Kinematics and Temperatures of Very Local ISM versus Warm Interstellar Gas}\label{sec:kinematics}

\begin{figure}
\plotone{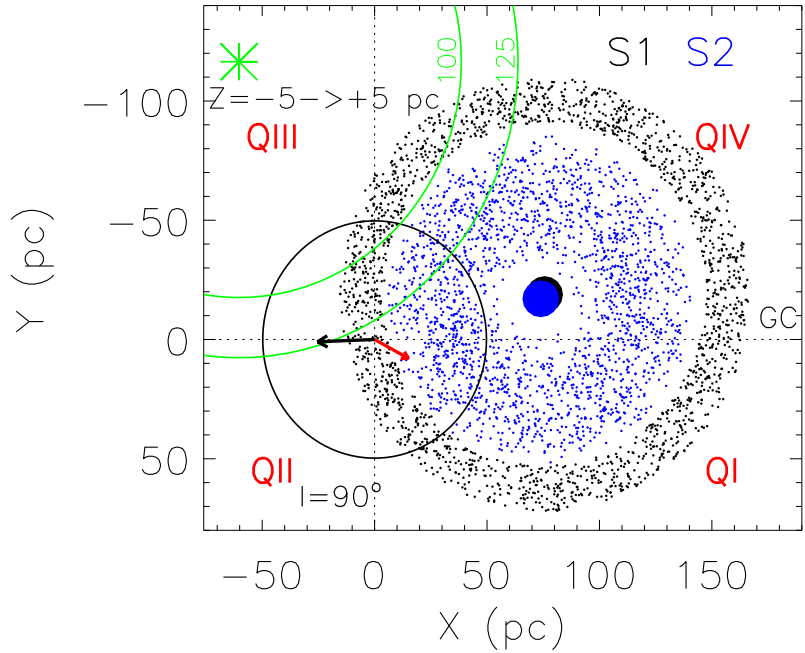}
\caption{A slice of the S1 (black dots) and S2 (blue dots) shells
within 5 pc of the galactic plane ($|$Z$| < 5$ pc) are plotted in x,y
coordinates, where the x-axis is directed towards the galactic center
(from \citep{Frisch:2008S1}).  Red and black arrows show the LSR
motions of the Sun and CLIC, respectively.  The black circle is 50 pc
from the Sun.  The two superimposed blue and black large dots show the
centers of the two shells.  The notation Q1--QIV indicate galactic
quadrants.  The green star shows the x,y position of the brightest
extreme ultraviolet (EUV) source in the sky, $\epsilon$ CMa, which is
located 15 pc below the galactic plane.  The two concentric green
circles show regions 100 pc and 125 pc equidistant from $\epsilon$
CMa.  The S1 and S2 shells are constructed using the shell parameters
in \citep{Wolleben:2007}.  }
\label{fig:shell}
\end{figure}

Using $Copernicus$, IUE, and optical data inside of the heliosphere
and towards nearby stars such as $\alpha$ Oph at 14 pc, Frisch
\citep{Frisch:1981} showed that the ISM inside and close to the
heliosphere has the kinematic and abundance properties expected for an
origin related to the Loop I superbubble.  The first spectrum of \lya\
fluorescence from interstellar \HI\ inside of the heliosphere,
acquired by $Copernicus$ during 1975 \citep{AdamsFrisch:1977}, yielded
the \HI\ velocity in the upwind direction of $\sim -24.7$ \kms\
(neglecting heliospheric acceleration and converting to the current
upstream direction \citep{Witte:2004,Frisch:2008S1}).  This \HI\
velocity projects to $\sim -21.1$ \kms\ in the $\alpha$ Oph direction,
and differs somewhat from the dominant cloud velocities known for that
direction of $ \sim -24 \pm 1$ \kms\ \citep{MarschallHobbs:1972}.  It
is now known that the \lya\ line backscattered emission has a
significant contribution from secondary \HI\ atoms, and also that the
LIC velocity observed inside of the heliosphere differs by $\sim 1$
\kms\ from the gas velocity towards the nearest star $\alpha$ Cen,
$\sim$50\deeg\ from the heliosphere nose \citep{LinskyWood:1996}, and
$\sim 3$ \kms\ from velocities of nearby gas in the upwind direction
towards 36 Oph \citep{Wood36Oph:2000}.  A more complete picture of the
kinematics and temperature structure of the LISM is now available.
The Sun is embedded in an ISM flow, the complex of local interstellar
clouds (CLIC),which has an upwind direction in the Local Standard of
Rest (LSR) directed towards the center of the S1 subshell of the Loop
I superbubble shell around the Scorpius-Centaurus Association
\citep{FrischYork:1986,Frisch:1995,FGW:2002,Wolleben:2007,RedfieldLinsky:2008}.
Fig. \ref{fig:shell} shows the S1 shell of Wolleben, found by fitting
1.4 GHz and 23 GHz polarization data, the solar apex motion, and the
bulk motion of the CLIC through the local standard of rest (LSR).  The
CLIC LSR upwind direction, $(\ell,b)\sim (358^\circ,-5^\circ)$
\citep{FGW:2002,FrischSlavin:2006book} is $\sim 10^\circ$ from the
center of the S1 shell at $(\ell,b)\sim (346^\circ \pm 5^\circ,3^\circ
\pm 5 ^\circ)$.  The CLIC kinematics is thus naturally explained by
the expansion of the S1 shell to the solar location
\citep{Frisch:1981}.  The expansion of Loop I has been modeled by
Frisch \citep{Frisch:1995,Frisch:1996}, and corresponds to an origin
during a star formation epoch $\sim 4 - 5$ Myrs ago.

Morphologically prominent shells such as the S1 shell are common
features in the ISM, often found between spiral arms where spurs are
seen.  Shell properties have been surveyed in the \HI\ 21-cm hyperfine
transition, revealing filamentary structures consisting partly of warm
neutral material (WNM).  Column densities for WNM are typically
\NHI$>10^{19}$ \cmtwo. WNM with column densities comparable to the
CLIC, \NHI$<10^{18.5}$ \cmtwo, or LIC, \NHI$\sim 10^{17.6}$ \cmtwo,
are not yet observed.  Zeeman data show that shell structures are
associated with magnetic fields of $\sim 6.4$ \microG\ or less.
Unfortunately, Zeeman-splitting data show that flux-freezing does not
occur in low density ISM, $n < 10^3$ \cc\ \citep{CrutcherHT:2003}, so
the magnetic field strength at the solar location can not be inferred
from the magnetic field in more distant portions of the S1 shell.

Turning back to the question ``Is the Sun embedded in a typical
interstellar cloud''.  I use the Arecibo Millennium Survey of the \HI\
21 cm line to define the meaning of ``typical''.  The Arecibo survey
is a complete and unbiased survey of warm and cold interstellar
clouds, as seen from the tropical Arecibo sky \citep{HTI}.  The
systematic fitting of Gaussian components to the emission profiles
revealed that 60\% of the ISM mass is contained in warm neutral
material (WNM), with median cloud column densities of $1.3 \times
10^{20}$ \cmtwo, compared to the lower median column density of the
cold neutral medium (CNM) of $5 \times 10^{19}$ \cmtwo.

\begin{figure}
\begin{center}
\plottwo{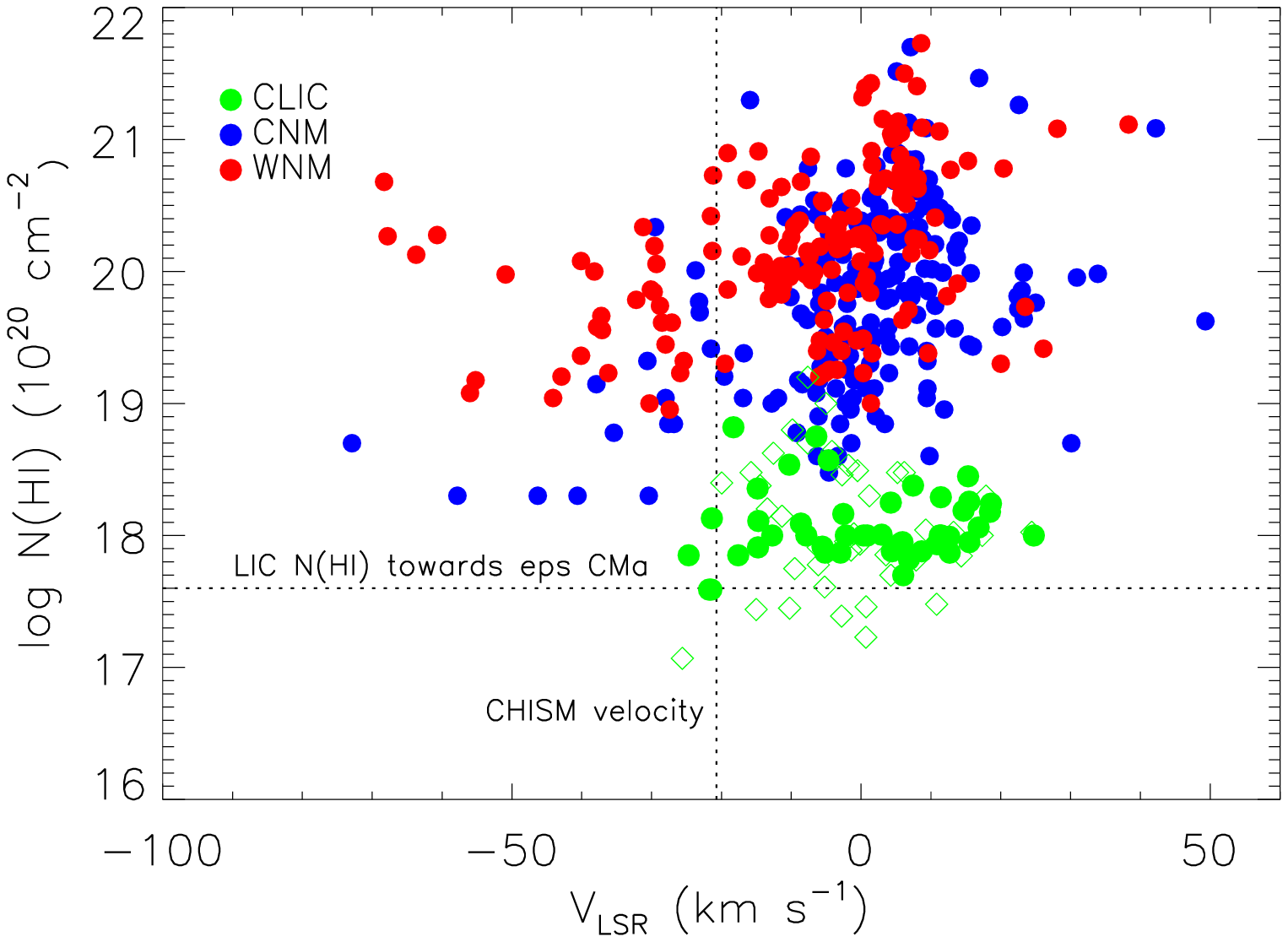}{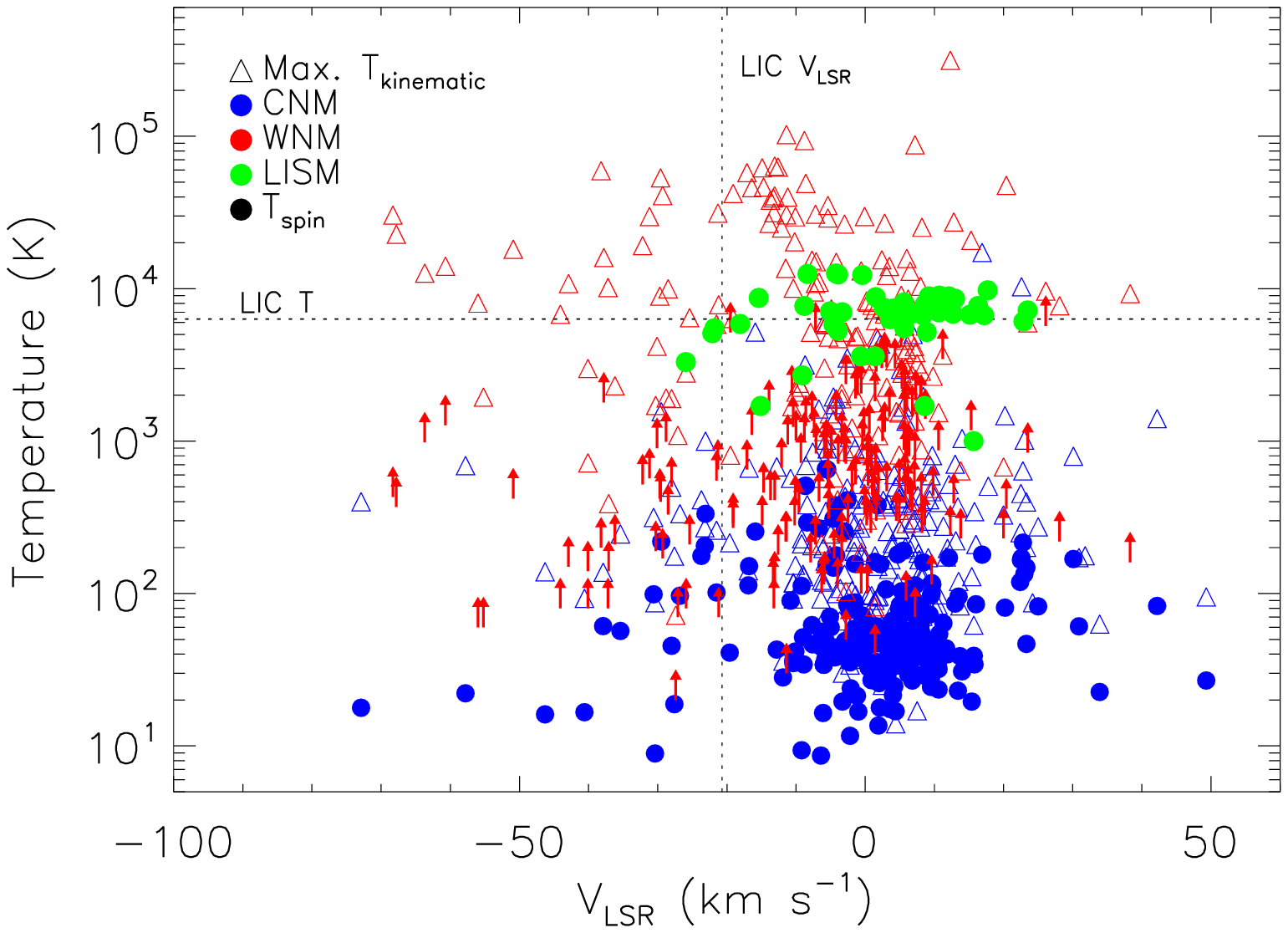}
\end{center}
\caption{The column densities \NHI, left, and \HI\ temperatures,
right, are plotted as a function of velocity for the WNM (red) and CNM
(blue) components in the Arecibo survey \citep{HTI} and for the CLIC.
The spin temperatures (red arrows) obtained for WNM by \citep{HTI} are
lower limits.  The CLIC components (green dots) are based on \CaII\
(diamonds, e.g. \citep{FGW:2002,Frisch:2008S1}), \HI, and \DI\ (dots,
\citep{RLIII,Woodetal:2005}). The CHISM LSR velocity is indicated,
based on the Standard solar apex motion. }
\label{fig:ams}       
\end{figure}

The kinematics of the Arecibo clouds can be used as a benchmark for
answering the above question as it applies to cloud kinematics.  In
Fig. \ref{fig:ams}, left, the kinematics of the CLIC cloudlets (LSR
velocities) are compared to the kinematics of the WNM and CNM.  For
the CLIC velocities I use \CaII\ and UV absorption line data such as
\DI\ \citep{FGW:2002,RLIII,Woodetal:2005}, and for plotting purposes
the ratios \CaII/\HI$=10^{-8}$ and \DI/\HI$=10^{-4.82}$.  It is
immediately apparent that the kinematics of CLIC clouds are comparable
to the global kinematics of both WNM and CNM clouds in the Arecibo
survey.

A second test using the Arecibo is also made.  The range of
temperatures for the WNM are shown in Fig. \ref{fig:ams}, right.  CLIC
temperatures from \citep{RLII} are also plotted, although there are
still poorly understood aspects of these temperatures (\S
\ref{sec:depletions}).  The Arecibo temperatures shown for WNM include
both spin temperatures (red arrows) and the kinetic temperatures (red
triangles) based on the FWHM of the fitted components.  For the WNM,
the kinetic and spin temperatures are upper and lower limits on the
thermal temperature, respectively, because turbulence is not removed,
and the true spin temperature is a function of a limit on the cloud
opacity \citep{HTI}.  The median kinetic temperatures for the WNM for
clouds for latitudes $b <30^\circ$ versus $b >30^\circ$ are,
respectively, 5,962 K, and 5,182 K.  The same ratio of low-to-high
latitude WNM temperatures is found for spin temperatures (1.15).  Low
and high latitude WNM median column densities are, respectively,
\NHI=$10^{20.68}$ \cmtwo\ and \NHI$=10^{19.98}$ \cmtwo.  Since the
CHISM temperatures is 6,300 K, it is within the WNM temperature range.
Fig. \ref{fig:ams}, right, shows that the temperatures of the CLIC
clouds (green dots) fall consistently between the upper and lower
limits on the WNM temperatures.  In addition, the CHISM temperature is
close to the median kinetic temperature of WNM components with $b <
30^\circ$.  The typical CLIC column densities, \NHI$<10^{18.5}$
\cmtwo, are below the range of detected WNM column densities.  Since
photoionization dominates the heating of the CHISM
\citep{SlavinFrisch:2008}, the higher low-latitude WNM temperatures
also suggest that radiation heating of the ISM is stronger at low
latitudes than high latitudes.

Based on the kinematical and temperature information in the Arecibo
Millennium Survey, the answer to the question posed above is again
'yes'.

\section{Partially Ionized  Gas and the Interstellar Radiation Field} \label{sec:pwim}

Two coupled attributes dominate the CHISM: it is partially ionized,
and it is low column density.  The first attribute follows from the
second in the presence of photons with energies $>13.6$ eV able to
ionize \HI.  The earliest observations of \HI\ and \HeI\ inside the
solar system found ratios of \HI/\HeI$\sim 6$
\citep{Ajelloetal:1987,WellerMeier:1981}, in contrast to EUV
observations of five white dwarf stars with average distances of 57 pc
and \HI/\HeI$\sim 14$ \citep{Frisch:1995}.  Cheng and Bruhweiler
\citep{ChengBruhweiler:1990} found that hot star radiation dominates H
ionization of the LISM but soft X-rays produced He ionization, and
therefore yielding higher ionization levels for He compared to H.
More recent studies show that He ionization is produced by EUV
emission from a conductive interface between the LIC and LB plasma,
white dwarf stars, and the low energy tail of the soft X-ray
background (\citep{SlavinFrisch:2008}, Slavin, Shelton, this volume).
The low H/He ratio found inside of the heliosphere thus becomes
evidence of the strong H filtration in heliosheath regions.

Ionized gas is a major component of the solar neighborhood.  FUSE
observations of ISM towards white dwarf stars within 70 pc find up to
$\sim 70$\% ionization levels, and electron densities in the range
0.025--0.25 \cc\ for stars with \NHI$=10^{18.8} - 10^{19.6}$ \cmtwo\
\citep{Lehneretal:2003}.  Hydrogen is $\sim 23$\% ionized at the
heliosphere, which is within the ionization range obtained by FUSE.
Radiative transfer models \citep{SlavinFrisch:2008} that predict the
heliosphere boundary conditions show that the CHISM electron densities
of $\sim 0.05-0.09$ \cc\ are similar to electron densities found by
FUSE, and also in the diffuse ionized gas sampled by pulsar dispersion
measures and \Halpha\ recombination lines.

The distribution of ionized gas near the LB is dominated by classic HII
regions around hot stars; the Wisconsin H-$\alpha$ Mapper (WHAM)
survey of the red \Halpha\ line shows these regions beautifully
\citep{WHAM:2003}.  However, recombination emission from low density
ionized gas carries more subtle information about partially ionized
regions such as the LIC.  Ionized gas in the solar vicinity fills
$\sim 20$\% of the disk and is contained in warm diffuse low density
regions with $n \sim 0.1$ \cc\ and $T \sim 10^4$ K. Ionization of this
gas is powered by O-stars, and requires transparent voids through
which the O-star radiation can propagate; the Local Bubble is such a
void.  A detailed comparison of \HI\ and \Halpha\ in a $\sim 120$
square-degree of sky showed that at least 30\% of the \Halpha\
emission is both spatially and kinematically associated with warm \HI\
21-cm features, many of which are filamentary
\citep{ReynoldsTufteHeiles:1995}. Some of this \HII\ is in regions
physically distinct from the \HI\ gas.  Ionization levels reach 40\%
for these low density, $\sim 0.2 - 0.3$ \cc, clouds.  The temperature
of diffuse ionized gas varies between 6,000 K and 9,000 K, with higher
temperatures at higher latitudes \citep{HaffnerReynoldsTufte:1999}.
This result follows from the temperature dependence of the \Halpha\
intensity of $\sim T^{-0.9}$, and $[$\NII$]$ and $[$\SII$]$ data.  The
CHISM temperature of 6300 K is within the range for the \Halpha\
clouds.

Is the diffuse \Halpha\ emission formed in partially ionized gas
similar to the LIC?  The answer to this is 'probably', however whether
or not the LIC radiation field is typical of diffuse gas is an open
question.  Observations of the \HeI\ 5876 A recombination line in the
diffuse ionized gas yield low levels of ionized He compared to H,
although the dominant O-star ionization source would predict higher
levels of He ionization.  Reynolds \citep{Reynolds:2004} compared
LIC radiative transfer model results with the partially ionized LIC
gas for four sightlines through diffuse ionized gas where the
forbidden 6300 A \OI\ line is measured.  These sightlines indicated H
ionization fractions of $>70$\%, compared to the LIC value of $\sim
23$\%. In addition, for these diffuse gas data, the ionization
fraction of He is 30\%--60\% of that of H, but the absolute He
ionization level is similar to the LIC.  Together these results
suggest that the radiation field at the LIC is harder than the diffuse
radiation field that maintains the warm ionized medium.  From the
relative H and He recombination lines, one thus might conclude that
the LIC is not typical of diffuse ionized gas.  However, the LIC
emission measure is $ EM \sim 0.003 $ cm$^{-6}$ pc, which is below the
WHAM sensitivity.  Radiative transfer models show that very low column
density clouds such as the LIC are transparent to H-ionizing
radiation, and such clouds may be invisible to WHAM.  The question as
to whether the relative ionizations of H and He in the LIC is typical
of ionized gas thus remains an open question, but the low LIC column
density probably explains the hardness of the local radiation field
compared to more distant regions.

Other properties of the interstellar radiation field that are
important for LIC ionization include the EUV and soft X-ray fluxes
\citep{SlavinFrisch:2008}.  Some doubt has been cast on the absolute
flux level of interstellar photons with energies $< 0.25$ keV because
of contamination of the soft X-ray background (SXRB) by heliospheric
emissions at energies $>0.4$ keV from charge-exchange between
interstellar neutrals and the solar wind (Shelton, Koutroumpa, this
volume). At 0.1 eV, LB emission has been modeled as contributing $\sim
50$\% of the flux \citep{HenleyShelton:2007}.  Clumping in the ISM may
change this picture, however, since a typical value of \NHI$\sim
10^{21}$ \cmtwo\ may include tiny cold clouds such as the \NHI$\sim
10^{18}$ \cmtwo\ structures that are completely opaque at low energies
\citep{StanimirovicHeiles:2005}.  If the X-ray emitting plasma
contains embedded clumps of ISM with significant opacity at $\sim 0.1
- 0.2$ keV, the energy dependence of the ISM opacity will be
significantly altered from that of a homogeneously distributed ISM
\citep{KahnJakobsen:1988}.  This effect will be significant for Loop I
X-ray emission, where embedded molecular clouds are found.  The
physical properties of the Local Bubble plasma need to be revisited by
including not only foreground emission from charge-exchange between
the solar wind and interstellar \HI, but also clumping in the ISM as
noted by \citep{KahnJakobsen:1988}.

There is one point that is not yet appreciated.  Since the hot star
$\epsilon$ CMa dominates the 13.6 eV radiation field in the solar
vicinity, and therefore the flux of H-ionizing photons, sightlines
through the third and fourth galactic quadrants (QIII, QIV),
\glong=180\deeg to \glong=360\deeg, will sample ISM with higher
ionization levels than sightlines through the first two galactic
quadrants \citep{Frisch:2008S1}.  This occurs because ISM associated
with the S1 shell structure is closer to $\epsilon$ CMa in QIII and
QIV, than in the first two galactic quadrants.  The relative locations
of the S1 shell and $\epsilon$ CMa are shown in Fig. \ref{fig:shell}.

\section{Chemical Composition of the ISM at the Sun } \label{sec:depletions}


The outstanding feature of warm low density interstellar clouds is
that the abundances of refractory elements such as Fe, Ti, and Ca, are
enhanced by an order of magnitude when compared to abundances in cold
clouds at $\sim 50$ K.  The enhanced abundances were originally
discovered for the \CaII\ line seen in high-velocity clouds
\citep{RoutlySpitzer:1952}, although the importance of the ionization
balance between \CaIII\ and \CaII, which favors \CaIII\ in warm
ionized gas, was not fully appreciated at that time. Enhanced
abundances are particularly strong for Ti, which is one of the first
elements to condense onto dust grains with $T_\mathrm{condensation}
\sim 1500$ K.  Column densities of \TiII\ can be directly compared
with \HI\ abundances without ionization corrections because \TiII\ and
\HI\ have similar ionization potentials.  Enhanced refractory element
abundances in warm gas at higher velocities has been modeled as due to
the destruction in shock fronts of refractory-laden interstellar dust
grains composed of silicates and/or carbonaceous material
(e.g. \citep{Slavinetal:2004}).  The CLIC gas shows such abundances,
requiring the CLIC grains to have been processed through shocks of
$\sim80$ \kms\ (Slavin, this volume, and \citep{Frischetal:1999}).
Refractory elements such as Mg, Si, Fe, and Ca are predominantly
singly ionized in the LIC, so that ionization corrections are required
to obtain accurate abundance information.  Ionization corrections are
generally not available for determining abundances of distant warm
gas; however the range of uncertainty in elemental abundances is large
enough that with or without ionization corrections, the CLIC gas has
typical abundances for low density clouds (e.g. \citep{Welty23:1999}).
The radiative transfer models provide accurate CHISM abundances that
are discussed by Slavin (this volume), and except for one sightline
CHISM abundances are typical for low density ISM.

There is only one sightline through the CLIC that shows a poorly
understood abundance pattern, and this is the sightline of $\alpha$
Oph that led to my original conclusion that the Loop I superbubble
shell has expanded to the solar location \citep{Frisch:1981}.  The
strongest observed \CaII\ line in the CLIC is towards $\alpha$ Oph,
where strong \TiII\ is also seen.  The star $\alpha$ Oph is 14 pc
from the Sun in the direction of the North Polar Spur, and the
interstellar gas in this sightline may be in the region where the S1
and S2 shells are in collision, so that
shock destruction of the grains is underway.

Two caveats must be attached to most determinations of elemental
abundances: (1) Common refractory elements tend to have FIP's $<13.6$
eV, so that ionization corrections are required to obtain accurate
abundances.  (2) Accurate \HI\ column densities are also required so
that abundances per H-atom can be calculated.  The first requirement
is seldom met, because cloud ionization data at best typically return
electron densities calculated based on either \MgI/\MgII\ or
\CIIstar/\CII\ ratios.  Total \HII\ column densities are not directly
measured and must be inferred.  Optimally, radiative transfer models
of each cloud could provide the same quality of results now available
for the LIC.  The second requirement is notoriously difficult to
achieve for \HI\ values relying on the heavily saturated \lya\ line.

\begin{figure}
\plottwo{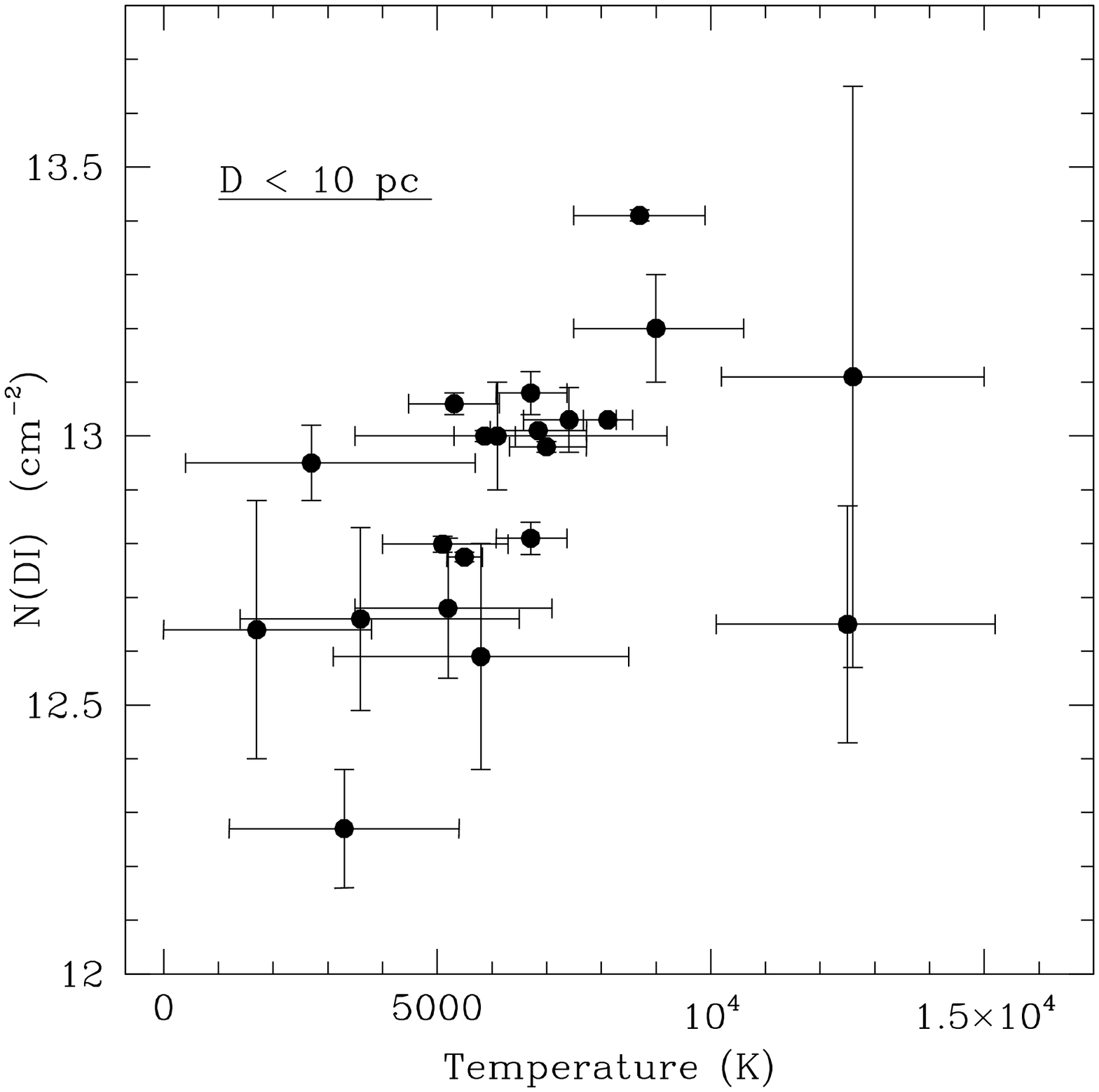}{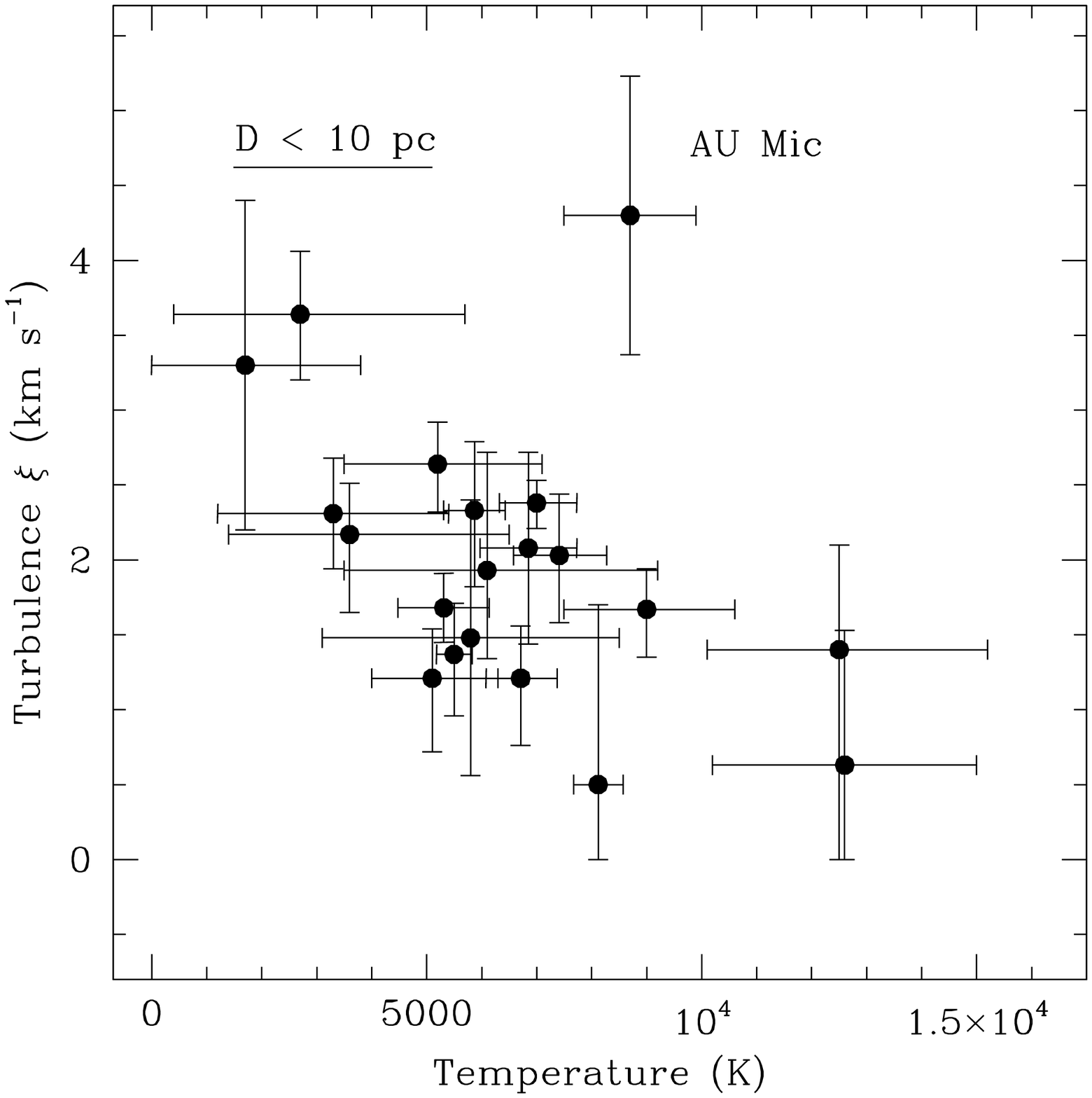}  
\caption[Cloud temperature and turbulence.]{\label{fig:tt} Temperature versus
column density $N$(DI) (left ) and temperature versus turbulence $\xi$
(right) for interstellar absorption components seen towards stars
within 10 pc of the Sun (data from \citep{RLIII}).  }
\label{fig:temp}       
\end{figure}

There are difficulties with extracting reliable column
densities from lines where thermal line-broadening dominates turbulent
line-broadening.  The Voigt profile used to determine the parameters
of absorption lines invokes the Doppler b-value ($b_\mathrm{D} \sim FWHM/1.7$):
\begin{equation}
b_\mathrm{D}(T,m)^2 = b_\mathrm{thermal}(T,m)^2 +
b_\mathrm{turbulent}^2,
\end{equation}
 where $b_\mathrm{turbulent}$ has no mass ($m$) or temperature ($T$)
dependence.  This assumption appears to break down for stars within 10
pc of the Sun, as is shown in Fig. \ref{fig:tt} (using data from
\citep{RLIII}) by the correlation between $N$(\DI) and temperature
$T$, and the anticorrelation between $b_\mathrm{thermal}(T,2)$ and
turbulence $\xi = b_\mathrm{turbulent}$.  No known cloud physics
explains a correlation between \DI\ and $T$ that is accompanied by an
anticorrelation between turbulent and thermal broadening.  One
explanation for this effect is that the assumption of isotropic
Maxwellian gas velocities and mass-independent turbulence breaks down
in a partially ionized low density ISM due to the coupling between
ions and magnetic fields.

The summary conclusion of this section is that CLIC and CHISM
abundances are similar to abundances in partially ionized gas.
Because of the uncertainties, this statement holds true when elemental
abundances are correctly compared to \HI+\HII, or \HI\ alone. The one
caveat on this statement is that \DI\ column densities for stars
within 10 pc show evidence of correlations that indicate the
line-broadening parameter is incorrectly defined.  The one sightline
that is not typical is $\alpha$ Oph, which may hold hidden clues about
colliding superbubbles near the Sun.

\section{Interstellar Magnetic Field at the Solar Location }\label{sec:Bfield}

The orientation, but not the polarity, of the interstellar magnetic
field (ISMF) at or near the heliosphere can be derived from optical
polarization vectors for nearby stars, $< 30$ pc.  This orientation
can then be compared with the local magnetic field direction derived
from the S1 shell low frequency radio continuum polarization (1.4 GHz,
Wolleben \citep{Wolleben:2007}, Frisch \citep{Frisch:2008S1}).  The
strongest optical polarizations are seen for stars located along the
ecliptic plane and with a peak in the polarization that is offset by
$\lambda \sim 40^\circ$ from the direction of the heliosphere nose
\citep{Frisch:2008S1}.  The orientation of the S1 shell magnetic field
in the heliosphere nose region agrees with the values obtained from
the optical polarization direction, to within the uncertainties, for
the Wolleben \citep{Wolleben:2007} angle parameter $B_\phi =
-42^\circ$.  The magnetic field at the position of the polarized stars
forms an angle of $\sim 65^\circ \pm 9^\circ$ with respect to the
ecliptic plane, and $\sim 55^\circ \pm 9^\circ$ with respect to the
galactic plane.  At the position of the \HeI\ inflow direction, the S1
shell configuration consistent with the polarization data gives a
magnetic field inclination of $\sim 73^\circ \pm 10^\circ$ with
respect to the ecliptic plane, and $\sim 44^\circ \pm 10^\circ$ with
respect to the galactic plane.  When the uncertainties on the upwind
directions of interstellar \HI\ and \HeI\ flowing into the heliosphere
are considered, then the offset angle between these two inflow
directions is $4.8^\circ \pm 0.6^\circ $, and these two upwind
directions define an angle of $55^\circ \pm 20^\circ$ with respect to
an ecliptic parallel
\citep{Frisch:2008S1,Moebiusetal:2004,Lallementetal:2005}.  When the
uncertainties are considered the \HI--\HeI\ offset angle, and the S1
shell direction that is consistent with the optical polarization data,
yield consistent ISMF orientations.

A non-zero angle between the ISMF direction and the inflowing ISM
velocity vector causes an asymmetric heliosphere, including a possible
tilt of $\sim 12^\circ$ between the heliosphere nose, as defined by
the maximum outer heliosheath plasma density, and the ISM velocity
(e.g. \citep{Ratkiewiczetal:1998,Lindeetal:1998,Ratkiewiczetal:2008}).
Is the S1 shell field orientation at the heliosphere consistent with
models of the known asymmetries of the heliosphere?  Pogorelov et
al. \citep{PogorelovOgino:2008} argue that large angles between the
upwind ISM and magnetic field directions are required to reproduce the
heliospheric asymmetry seen by the Voyager 1 and Voyager 2, which
encountered different termination shock distances at $\sim 94$ au and
$\sim 84$ au \citep{Stone:2008}.  A magnetic field direction tilted by
$\sim 60^\circ$ with respect to the ecliptic plane reproduces the
offset angle between \HI-\HeI, but not the heliospheric asymmetry seen
by the Voyager satellites (Pogorelov et
al. \citep{PogorelovOgino:2008}).  Opher \citep{Opher:2008} reports
that an interstellar field direction inclined by $\sim 60^\circ -
90^\circ$ with respect to the galactic plane reproduces the Voyager
results, including particle streaming in the outer heliosheath.
Ratkiewicz et al. \citep{Ratkiewiczetal:2008} find that an ISMF
directed towards galactic coordinates \glong$=217^\circ \pm 14^\circ$,
\glat$=-50^\circ \pm 9^\circ$ explains the position of the \lya\
maximum observed by the Voyager spacecraft in the outer heliosphere.
These models, the interstellar polarization data, and the S1 shell
predictions of the ISMF direction at the heliosphere agree to within
the large uncertainties remaining in this problem.

\begin{figure}
\plotone{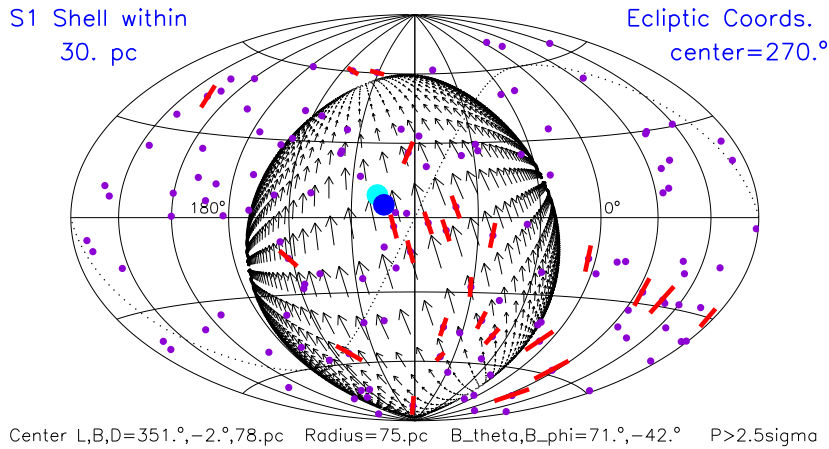}
\caption{ The magnetic field associated with the parts of the S1 shell
within 30 pc is plotted in ecliptic coordinates for an aitoff
projection.  The parameters for the S1 shell given in
\citep{Wolleben:2007} have been varied within the range of allowed
uncertainties to yield the best match to polarization data towards
stars in the heliosphere nose region (indicated by the dark blue
dots).  The dark and light blue dots show the inflow directions of
interstellar \HeI\ and \HI\ into the heliosphere.  The S1 subshell
parameters used in the above figures correspond to a shell center at
(\glong,\glat)=$(351^\circ,-2^\circ)$ and a distance of 78 pc away, a shell radius of 75
pc, and magnetic field angles $B_\theta =71^\circ$ and $B_\phi=
-42^\circ$.  The dots show stars within 50 pc with polarization data,
and the red bars show polarization vectors for stars where
polarizations are larger than $ 2.5 \sigma$
\citep{Tinbergen:1982,Piirola:1977,Frisch:2007cmb}.
\label{fig:shellB} }
\end{figure}

If magnetic and thermal pressures in the CHISM are approximately
equal, then the CHISM field strength is $ \sim 2.7 $ \microG\
\citep{SlavinFrisch:2008}.  The polarity of the CHISM field is a more
difficult question, and can presently only be inferred from the
polarity of the nearby global magnetic field.  The global magnetic
field direction in the solar vicinity is directed towards \glong$\sim
80^\circ$ \citep{Han:2006}.  For a classical expanding superbubble
model \citep{MacLowMcCray:1988}, where a shock front sweeps up
interstellar material and compresses magnetic field lines in the
expanding shell, the S1 shell expansion would have preserved the
global field polarity so that the S1 shell field direction at the Sun
is directed from the south to north.  With this polarity, the
direction of the interstellar magnetic field at the heliosphere nose
is shown in Fig. \ref{fig:shellB} (see \citep{Frisch:2008S1} for the
field direction in galactic coordinates).  This model neglects
possible additional rotation of the field direction, such as may arise
from coupling between the ISMF and $\sim 18^\circ$ tilt of the plane
of Goulds Belt with respect to the galactic plane.

Based on the similar magnetic field directions obtained from the S1
shell magnetic field at the heliosphere and the polarization of light
for stars close to the Sun, the magnetic field in the CLIC and CHISM
is typical.  The field strength inferred from pressure equilibrium in the
CHISM, $\sim 2.7$ \muG, is typical of field strengths found from
Zeeman splitting of the 21-cm line for WNM in the Arecibo survey, and
$\sim 50$\% larger than the large-scale ordered magnetic component
inferred from pulsar data \citep{RandKulkarni:1989}.  Given that there
is no evidence that the interstellar magnetic field at the heliosphere
is anomalous, again the answer to the posed question is 'yes'.

\section{Conclusions}

By all of the standard measures of interstellar clouds, such as
temperature, velocity, composition, ionization, and magnetic field,
the interstellar gas inside of the heliosphere and in the LIC are
typical of warm partially ionized gas seen elsewhere in the
neighborhood of the Sun.  Unfortunately clouds with low LIC-like
column densities are not yet observable in either \HI\ 21-cm or
\Halpha\ recombination lines, so that clouds with hard radiation
fields similar to the LIC can only be idenitified through ultraviolet
absorption lines.

The association of the LISM and LIC gas with the expanding S1
supersubble shell, and possibly the S2 shell, naturally explains the
kinematics of ISM within $\sim 30$ pc.  Furthermore, the S1 shell
structure leads to specific predictions about the relative ionizations
of different parts of the shell due to proximity to $\epsilon$ CMa and
other nearby hot stars \citep{FrischChoi:2008}.  The S1 shell also
predicts a direction of the interstellar magnetic field at the
heliosphere that is consistent with observations of optical
polarizations towards nearby stars.

It is evident that the answer to the question posed by the title of
this paper is 'yes', so this paper will close with a more difficult
question posed years ago by Eugene Parker: "What is an interstellar
cloud". Originally clouds like the LIC and other LISM clouds were
named "intercloud medium".  The LIC column density towards Sirius,
$\alpha$ CMa, suggests the Sun has entered the LIC within the past few
thousand years (Frisch \citep{Frisch:1994}), while the velocity
discrepancy between interstellar \HeI\ inside of the heliosphere and
ISM in the upwind direction towards $\alpha$ Cen and 36 Oph suggests
the Sun is at the edge of the LIC (Lallement et
al. \citep{Lallementetal:1995}, Wood et al. \citep{Wood36Oph:2000}).
Are there two separate clouds adjacent to the heliosphere?  Or instead
are we crossing a pocket of microturbulence with scale sizes of $\sim
0.02$ pc?  What is a cloud anyway?  The LIC is $\sim 10^{21}$
orders-of-magnitude less dense than the terrestrial atmosphere.


\begin{acknowledgements}
The author would like to thank NASA for support
through grants NAG5-13107, NNG06GE33G, and through the IBEX mission.
My collaborator Jon Slavin has made important contributions to much of
this work.  The author would also like to thank the International
Space Sciences Institute for sponsoring the workshop "From the Outer
Heliosphere to the Local Bubble: Comparison of New Observations with
Theory".

\end{acknowledgements}

\newcommand{\bibfont}{\footnotesize}
\bibliographystyle{plain}


\end{document}